\documentclass[prl,aps,reprint,twocolumn,showpacs]{revtex4-1}
\usepackage{graphicx}
\usepackage{amsmath}
\usepackage{amsfonts}
\usepackage{amssymb}

\usepackage{epstopdf}
\usepackage[usenames,dvipsnames]{xcolor}

\renewcommand{\imath}{i}

\begin{document}
	\title{FPU physics with nanomechanical graphene resonators: intrinsic relaxation and thermalization from flexural mode coupling}
	\author{Daniel Midtvedt}
	\email{midtvedt@chalmers.se}
	\affiliation{Department of Applied Physics, Chalmers University of Technology, 
					SE-412 96 G\"{o}teborg, Sweden}
	\author{Zenan Qi}
	\affiliation{Department of Mechanical Engineering, Boston University, Boston, MA 02215, USA}
	\author{Alexander Croy}
	\affiliation{Department of Applied Physics, Chalmers University of Technology, 
					SE-412 96 G\"{o}teborg, Sweden}
	\author{Harold S. Park}
	\affiliation{Department of Mechanical Engineering, Boston University, Boston, MA 02215, USA}
	\author{Andreas Isacsson}
	\affiliation{Department of Applied Physics, Chalmers University of Technology, 
					SE-412 96 G\"{o}teborg, Sweden}

	\date{\today}

\begin{abstract}
Thermalization in nonlinear systems is a central concept in statistical
mechanics and has been extensively studied theoretically since the seminal work
of Fermi, Pasta and Ulam (FPU). Using molecular dynamics and continuum modeling
of a ring-down setup, we show that thermalization due to nonlinear mode coupling
intrinsically limits the quality factor of nanomechanical graphene drums and
turns them into potential test beds for FPU physics.  We find the thermalization rate $\Gamma$ to be independent of radius and scaling as  $\Gamma\sim T^*/\epsilon_{{\rm pre}}^2$, where $T^*$ and $\epsilon_{{\rm pre}}$ are effective resonator temperature and prestrain. 
\end{abstract}

	\maketitle

Advances in fabrication techniques enable production and
characterization of one and two dimensional nanoscale mechanical resonators \cite{buza+07,erle+08,chro+09,eimo+11}. In particular, carbon-based resonators are considered to be promising for many applications due to their low mass and high quality factors (Q-factors) \cite{osha+12,bail+11,zaba+10,eopa+11}. It is also known that these systems display strongly nonlinear behavior \cite{atis+08,eimo+11}, which makes them interesting for investigations of nonlinear dynamics.
The nonlinearities lead to a coupling between the vibrational modes \cite{mavi+13,ermi+13}. This coupling  allows for intermodal energy-transfer, which facilitates the redistribution of energy initially localized in a single mode. 

In this respect, the mode-coupling provides a dissipation channel for the fundamental mode (FM) dynamics. In contrast to other dissipation mechanisms previously studied in nanomechanical resonators \cite{liro00,crli01,wil08,rebl+09,crmi+12,immo+13}, this is a fundamental intrinsic mechanism and therefore constitutes a lower limit on the relaxation rate of the FM. At finite temperatures, the effect of the mode couplings will be two-fold. First, they give rise to fluctuations in the resonator strain leading to dephasing or spectral broadening of the resonator \cite{basa+12}. Second, as we show in this Letter, they allow for energy redistribution among the modes. To distinguish the two effects we consider a ring-down setup, where the total energy of the resonator is conserved and the evolution of the spectral distribution of energy is monitored. This allows us to access the dynamics of the FM energy.

The process of thermalization in a system of nonlinearly coupled oscillators was originally considered by Fermi, Pasta and Ulam (FPU) in their famous computer experiment in 1955 \cite{FPU+55,ca+05}, and spawned an impressive amount of research that eventually resulted in the development of chaos theory \cite{izch+66} and the discovery of solitons \cite{zakr+65}. For the FPU problem, it is known that above a certain critical energy density, energy initially fed into the FM is quickly redistributed among all modes, and the system approaches a thermal state. This threshold is connected to the onset of widespread chaos in the mode dynamics \cite{izch+66,ca+05} and the stability of localized modes ("q-breathers") \cite{pefl+07,fliv+05}. In recent years, the consensus has been reached that the main features observed in the FPU problem are not specific to the original model Hamiltonian \cite{fuma+82,bapo+08}. A natural question, which is still under debate \cite{beca+08}, is whether those features can be observed in a physical system. For this to be possible two requirements need to be fulfilled: first, the nonlinearity has to be sufficiently strong to allow for appreciable coupling between resonator modes already at low energies. Second, the time-scale of energy dissipation to the environment must be long compared to that for thermalization due to the mode coupling. We propose that nanomembrane resonators can be used to test the persistence of the FPU phenomena in the thermodynamic limit. 

While mode-coupling is present in any crystalline membrane, we here focus on graphene due to the many realizations of membrane resonators using this material. To study the intrinsic loss mechanisms in such systems, we utilized classical molecular dynamics (MD) simulations to systematically investigate the free vibrations of pristine circular graphene monolayers with varying radius, pre-strain, temperature and excitation energy. The MD simulations were performed using LAMMPS \cite{plimptonLAMMPS, plimptonJCP1995}. The covalent bonds between carbon atoms were modeled by the AIREBO potential \cite{stuartJCP2000,zhaoNL2009, qiNS2012, qiNANO2010}.

After an initial relaxation stage, the graphene monolayer was strained and fixed at its edges. The system was then equilibrated at a specific temperature within the canonical ensemble (NVT) for 10 ps.  Thereafter, the monolayer was actuated by assigning an initial velocity profile in the out of plane direction corresponding to the fundamental mode shape of the resonator.  After that point, the system was allowed to vibrate freely in the micro canonical (NVE, or energy conserving) ensemble for 5000 ps with a time step of 1 fs. The entire simulation was divided into 10 time windows of 500 ps each; the first 100 ps of each time window were used for further analysis.  Specifically, total kinetic energy and velocity history of each atom were recorded to calculate the total kinetic energy spectrum.  We studied four temperatures ($T$ = $\sim$0, 50, 100 and 300 K), four radii ($R$ = 5, 7, 9 and 11 nm) and four initial excitation energies ($v_{{\rm max}}$ = 2, 5, 10 and 15~\AA/ps). Additionally, the pre-strain $\epsilon_{\rm pre}$ of the structures was varied.

\begin{figure}
		\centering
		\includegraphics[width=1\columnwidth]{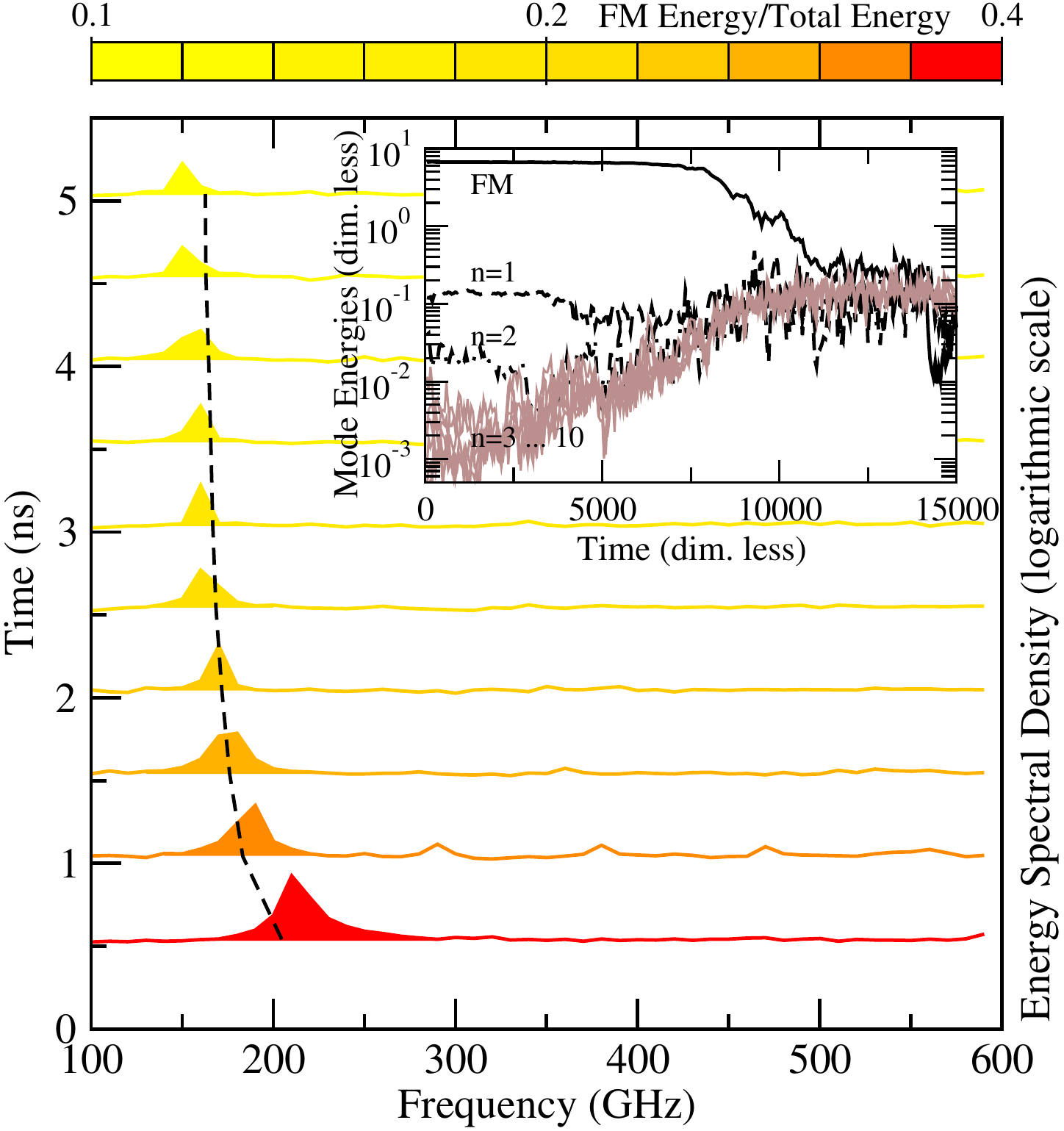}\hfill
		
\caption{Time evolution of the energy spectral density of a circular graphene sheet obtained by MD simulations for the out-of-plane motion. Parameters are given in the text. Note the decay of the FM energy which is indicated by the filled area under the curve, as well as the shift in the FM frequency. The dashed black line is the FM frequency as estimated from the CM model. The inset shows the evolution of the individual mode energies obtained from the CM model in logarithmic scale.}\label{fig:Spectrogram}
	\end{figure}

In Fig.\ \ref{fig:Spectrogram}, the time evolution of the kinetic energy spectral density of a circular graphene sheet of radius $5$ nm, $\epsilon_{{\rm pre}}=0.5$\% kept at a temperature of $300$ K is reported up to a time of $5$ ns. Initially, the FM is excited by a velocity of $10$Ê~\AA/ps. 

The spectral density of the kinetic energy in Fig.\ \ref{fig:Spectrogram} corresponds to the frequency distribution of the contribution to the kinetic energy from the out-of-plane motion. The in-plane contribution is negligible in the frequency range relevant for flexural vibrations. The area under a peak gives the kinetic energy of that mode. The prominent peak initially located around $\sim 200$ GHz corresponds to the FM, and its energy decreases continuously during the time evolution. Simultaneously, the frequency of the FM decreases. As the FM frequency is energy-dependent, this suggests that energy is redistributed among normal modes in the system.

This redistribution is known to occur in the problem of coupled Duffing oscillators. In fact, in a continuum mechanics (CM) approach \cite{ermi+13,atis+08} the present system is also described by a system of coupled Duffing oscillators. In dimensionless form, the equations of motion for the mode amplitudes $q_n$ are written as (see supplement)

	 %can we get this for R>= 9nm?

\begin{equation}
\label{eq:modecoup}
\partial^2_{\tau} q_n + \tilde{\omega}_n^2 q_n + \sum_{ijk} W_{ij;kn} q_i q_j q_k = 0.
\end{equation}
The coupling matrix $W_{ij;kn}$ depends only on the geometry of the resonator. For drum geometries it is, in contrast to the FPU case \cite{lili+08}, a dense matrix, with permutation symmetry in the indices $i \leftrightarrow j$ and $k \leftrightarrow n$ \cite{ermi+13}.
The dimensionless time and energy are related to the physical units through
\begin{equation}
\tau=\frac{\sqrt{\epsilon_{{\rm pre}}} c_{{\rm L}}}{R} t, \;  \tilde{E}=\frac{E}{2\pi \epsilon_{{\rm pre}}^2 c_L^2 \rho_{{\rm G}} R^2},
\end{equation}
where $c_{{\rm L}}=\sqrt{Y_{{\rm 2D}}/\rho_{{\rm G}}}$ is the longitudinal speed of sound in graphene, with $Y_{{\rm 2D}}\approx 350$ N/m being the two dimensional Young's modulus of graphene \cite{lewe+08} and $\rho_{{\rm G}}=0.76$ ${\rm mg/m^2}$ the graphene mass density. In deriving \eqref{eq:modecoup} we have further scaled the radial coordinate $r$ and the vertical displacement $w$ according to $\tilde{r}=r/R$, $\tilde{w}=(1/2) w/(\epsilon_{{\rm pre}}^2 R)$. The linear frequencies are given by the zeros $\xi_{0,n}$ of the zeroth order Bessel function, $\tilde{\omega}_n=\xi_{0,n}$.

The two models (CM and MD) complement each other. The MD is derived from an atomistic approach, but is computationally heavy and is restricted to relatively small systems. The CM equations are derived assuming negligible bending rigidity and long wavelength deformations, but require less computational power and further allow to predict scaling behaviors for various physical parameters. Since the CM model is a long wavelength approximation, only the low-frequency modes can be accurately described within the model. Additionally, in the CM-simulations presented here only radially symmetric modes are considered, as the interaction term in Eq. \eqref{eq:modecoup} conserves the radial symmetry of the fundamental mode.

The dimensionless and parameter free form of the CM equations implies that the dynamics is completely determined by the initial conditions. Further, strong mixing in phase space causes the distribution functions of the modes to decouple. The dynamics of the system is then described by the total dimensionless energy $\tilde{E}$ and the ratio of applied FM energy $\tilde{E}_0$ to total energy, $\eta=\tilde{E}_0/\tilde{E}$. 

There will also be a dependence on the number of degrees of freedom in the system, set by the number of atoms. In the CM model, this is introduced artificially through the number of modes $N$ that are considered in the simulations. This number is always much smaller than the number of atoms in the physical system.
	\begin{figure*}
		\centering
		\includegraphics[width=2.1\columnwidth]{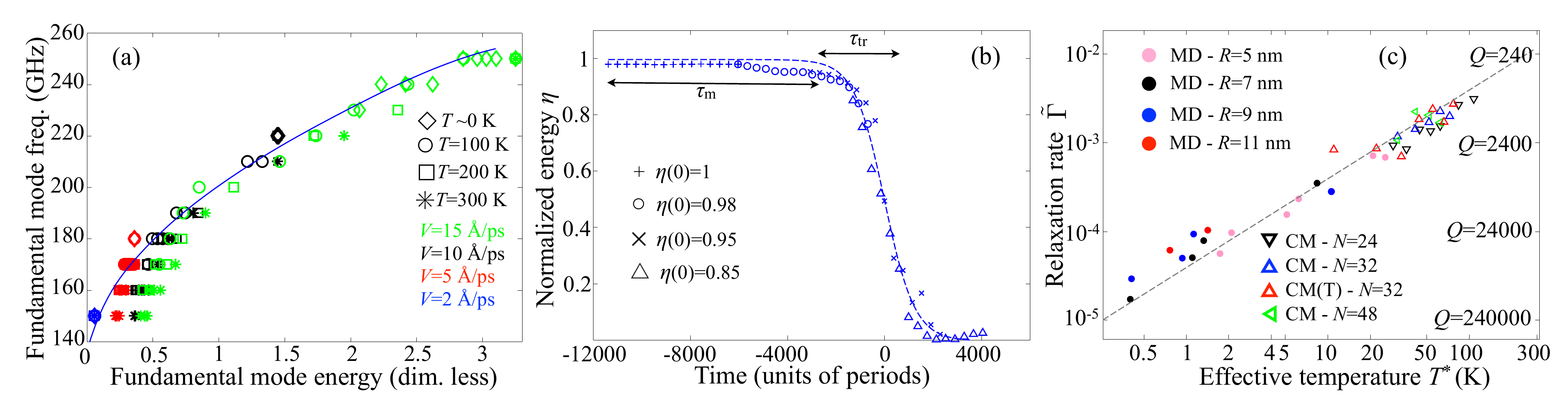}\hfill
		%\includegraphics[height=4.9cm]{Fig2b.pdf}\hfill
		%includegraphics[height=4.6cm]{Fig2cv2.pdf}\hfill
\caption{(a) Fundamental mode frequency as a function of mode energy. Symbols correspond to MD results, while the full line is the curve predicted from CM. The finite length of the time window used to calculate the frequency spectrum limits the frequency resolution. This is represented by the size of the symbols. (b) Simulation of Eq.\ \eqref{eq:modecoup} for a fixed total energy, but with varying initial FM energy. By shifting time, the curves align. The solid line is a fitted sigmoid function used to extract the transition time $\tau_{{\rm tr}}$. (c) Extracted rates $\Gamma=\tau_{{\rm tr}}^{-1}$ from MD simulations (filled symbols) and CM (open symbols) for fixed initial value of $\eta=1/2$ and varying total energy, reported as a function of the effective temperature $T^*$ for fixed $\epsilon_{{\rm pre}}=0.2$ \% according to Eq.\ \eqref{eq:dimE}. The dashed line corresponds to a linear scaling with $T^*$.}\label{fig:CMmodel1} %Error bars for resolution of STFT? Connect to Q-factors df=10GHz %Error in dim less E for MD?
	\end{figure*}

The dimensionless energy can be written in terms of the average energy per atom $\kappa$ as
\begin{equation}
\label{eq:dimE}
\tilde{E}=\frac{1}{2\epsilon_{{\rm pre}}^2}\frac{\kappa}{m_{{\rm c}}c_{{ \rm L}}^2 },
\end{equation}
where $m_{{\rm c}}$ is the carbon atom mass. For a system of uncoupled harmonic oscillators in equilibrium, $\kappa=k_{{\rm B}} T$. We study a non equilibrium situation, but an effective temperature may be defined by considering the energy not residing in the FM, $k_{{\rm B}} T^* \equiv (E-E_0)/N$ where $N$ is the number of degrees of freedom. The relation \eqref{eq:dimE} shows the correspondence between temperature and strain in this system. The importance of the mode coupling is determined by the thermal fluctuations of the membrane. These may be enhanced either by increasing the temperature, or by decreasing the strain. 

The shift of the FM frequency in Fig.\  \ref{fig:Spectrogram} can be reproduced in the CM model. In Fig.\ \ref{fig:CMmodel1}(a), the energy dependence of the FM frequency is reported for the MD simulations (symbols), together with the predicted curve from the CM model. The frequency shift is a result of nonlinearities, and the overall agreement indicates that the nonlinearities are well described within the CM model.

Next we consider the dynamics of the FM energy. The inset of Fig.\ \ref{fig:Spectrogram} shows the temporal evolution of the individual mode energies in the CM model, when initially all energy is fed into the fundamental mode. Note the appearance of an initial metastable state, with nearly all energy localized in the fundamental mode. Among modes with mode number $>3$, the energy is always equipartitioned, indicating that the assumption of strong mixing is valid. These modes define an instantaneous effective temperature of the system, which due to the redistribution increases in time. This effect is more pronounced in the CM-simulations due to the mode number cut-off.

In Fig.\ \ref{fig:CMmodel1}(b) the normalized FM energy $\eta$ is monitored as a function of time, by calculating the mode dynamics from Eq.\ \eqref{eq:modecoup} for a system of $N=32$ modes for a fixed total energy $\tilde{E}=3.25$. The degree of initial excitation over the thermal background, i.e. $\eta$, is varied in the simulations. Shifting the time, so that $\eta=1/2$ at $t=0$, the curves align. This indicates that the important parameters are the total energy and $\eta$. Further, two separate time-scales associated with the non-equilibrium dynamics of the system can be identified; initially, when most energy is in the FM, the system relaxes to a long-lived metastable state with energy dependent life-time $\tau_{{\rm m}}$. In this region, the system is sensitive to fluctuations and the relaxation is therefore strongly temperature dependent. Thereafter the system undergoes a sharp transition to an equipartition region during a time $\tau_{{\rm tr}} = \Gamma^{-1}$. 

The MD simulations are best suited to investigate the shortest of these timescales, the transition time $\tau_{{\rm tr}}$. We performed MD simulations at various total energies with initial data chosen such that the system was initially in the transition stage. The velocity applied to the FM was chosen such that $\eta\approx 1/2$ at the beginning of each simulation. An estimate of the inverse transition time is then given by the (dimensionless) time derivative of the ratio between fundamental and total energy, $\Gamma=\partial_{\tau} \eta=\xi_{0,0}\partial_t \tilde{E}_0/\omega_0 \tilde{E}$. This quantity may be considered as an "instantaneous Q-factor" of the resonator. The Q-factors reported here are evaluated at $\eta \approx 1/2$, where the relaxation rate is maximal. If temperature and excitation velocity are changed independently the ratio of fundamental to total energy will also change, which will obscure the scaling of the transition time-scale with energy. To avoid this, the total energy was tuned by changing the pre-strain, employing the duality between temperature and strain apparent from Eq.\ \eqref{eq:dimE}. The pre-strains of the structures used in the MD simulations were determined by considering the number of atoms in the resonators and the radius, using further the known value for the graphene bond length obtained from the AIREBO potential ($1.396$ ~\AA\ \cite{stuartJCP2000}). 

The results from the MD simulations (filled symbols) and CM model (open symbols) are compared in Fig.\ \ref{fig:CMmodel1}(c). There is no dependence of the dimensionless relaxation rate on the size of the drum. This is consistent with the observation that the dimensionless energy in Eq.\ \eqref{eq:dimE} does not contain any length scale. The transition rate is linear in energy for both models, i.e. $\Gamma \propto T^*/\epsilon_{\rm pre}^2$.

Note that previous studies on CVD (Chemical Vapor Deposition) grown resonators show a power law dependence of the Q-factor on radius \cite{bail+11}, in contrast to what is reported here. However, for CVD grown graphene grain boundaries will introduce an additional length scale that determines the mode coupling \cite{qiNS2012}. 
	
The existence of an initial metastable state for initial conditions far from equilibrium is a well known feature of the FPU-problem \cite{beca+08,fuma+82,ca+05}, and occurs also in other nonlinear lattices \cite{fuma+82} (see supplement). The metastable state is strongly localized in mode space, and is related to so called q-breathers \cite{fliv+05,pefl+07}. The metastable state is obtained when the energy in the FM is much larger than the thermal energy. External losses and thermal noise are expected to destroy this state, but traces of it may be found by considering the relaxation as a function of excitation energy. Numerical evidence on the FPU-chain suggests a stretched exponential dependence of the life-time with excitation energy, $\tau_{\rm m}\eqsim \exp\left(\tilde{E}^{-\alpha}\right)$ \cite{bega+84,beca+08}. 

\begin{figure}
		\centering
		\includegraphics[width=.95\columnwidth]{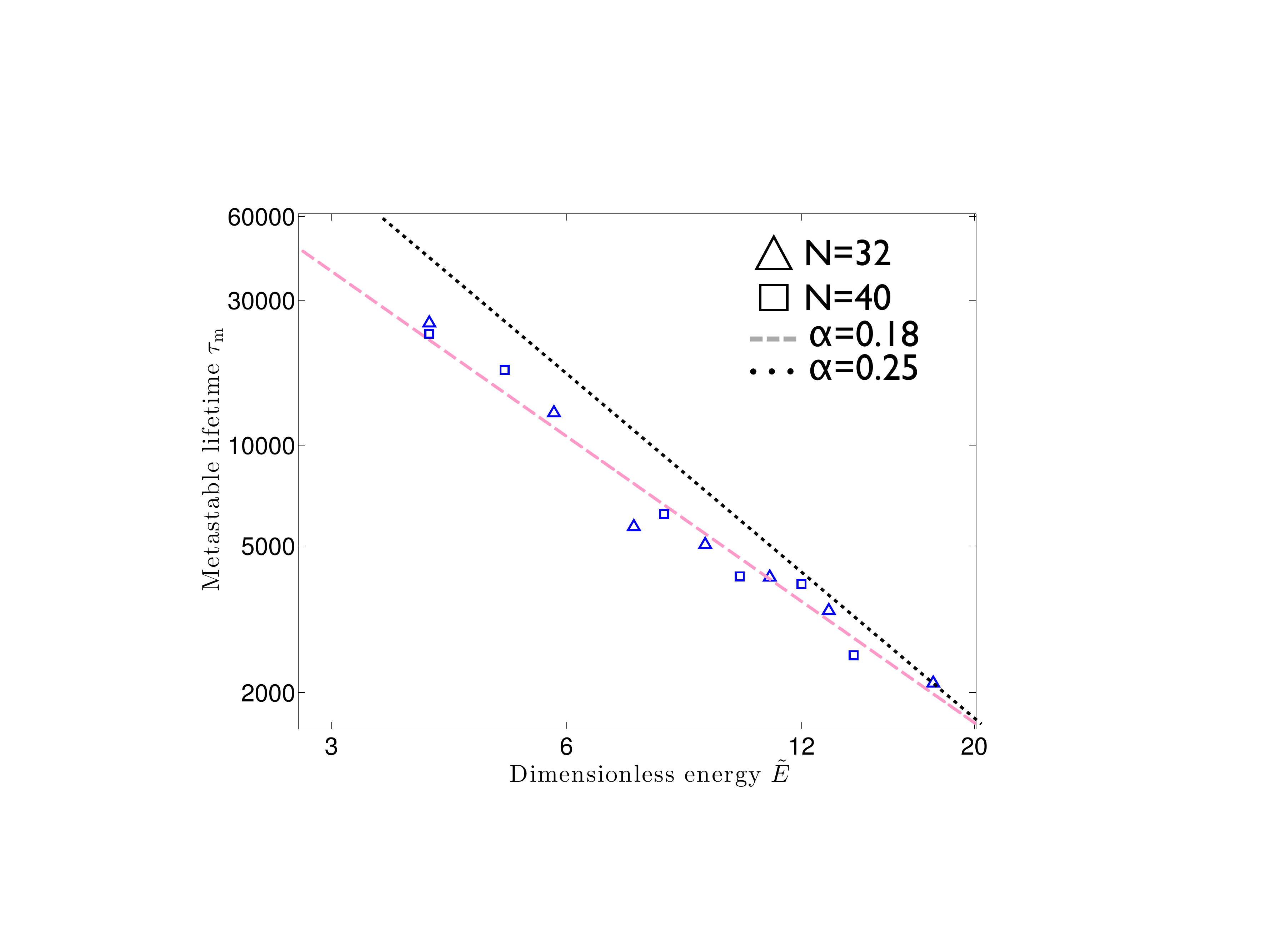}\hfill
		
\caption{Dependence of the metastable lifetime $\tau_{\rm m}$ on system energy, for the CM model with 32 modes (triangles) and 40 modes (squares). The lines corresponds to stretched exponentials of the form $\tau_{\rm m}\eqsim \exp\left(\tilde{E}^{-\alpha}\right)$. The dashed line corresponds to $\alpha=0.18$, the dotted corresponds to $\alpha=0.25$.}\label{CM:tau_m} 
\end{figure}

We probe the metastable state by initially feeding all energy into the FM, and monitoring the evolution of the FM energy. As the time-scales for the decay of this state is very long compared to the FM oscillations, this state is most readily investigated using the CM model. For the present model, we obtain an exponent $\alpha\approx 0.18$ from CM simulations with $N=32$ and $N=40$ modes (see Fig.\ \ref{CM:tau_m}). The life-time of the metastable state does not show a dependence on the number of modes. Note that the exponent $\alpha$ is model dependent \cite{pela+89}, e.g. for the FPU-$\beta$-problem $\alpha=0.25$ has been reported \cite{begi+04}.

The energy is strongly localized to the FM during the metastable state. The localization of the metastable state in mode space depends on the frequency spectrum. For the drum resonator, the frequencies can be approximated by $\omega_{n,{\rm drum}}\approx \xi_{0,0} + n\pi$. In comparison, for small wave numbers the frequencies of the FPU chain are given by $\omega_{n,{\rm FPU}}\approx n\pi$ \cite{beiz+05}. Low frequencies of the FPU chain are almost resonant, and so the formation of a q-breather will consist of a cascade of modes being excited. The present system, being far from resonant, displays a strong localization of energy.

The CM model allows us to make quantitative predictions for the relaxation rate due to energy redistribution in resonators of arbitrary size and tension. We take as an example a drum of arbitrary radius, $\epsilon_{{\rm pre}}=0.2\%$ and $T=4$ K. The relaxation rate can be read off from Fig.\ \ref{fig:CMmodel1}(c), giving a transition time of approximately $24000$ oscillations, decreasing to $\sim200$ oscillations at 300 K. For this relaxation to be observable, the Q-factor arising from external losses must exceed these values.

In summary, the redistribution of energy in graphene resonators due to nonlinear mode coupling has been investigated numerically in a ring-down setup. This coupling is a limiting factor for the stability of excitations of individual modes. At low temperatures, we find evidence for a state akin to the metastable state found in the FPU problem. After a time increasing exponentially with inverse excitation energy, the system relaxes toward its equilibrium with an energy dependent rate. 
%The experimental verification of this metastable state would constitute strong evidence for the importance of energy redistribution due to mode coupling in graphene nano resonators.

The rate of relaxation from a strong non-equilibrium situation with all energy in the FM to a situation close to equipartition depends only on the dimensionless energy of the resonator, which in turn depends on temperature and strain through the ratio $T^*/\epsilon_{{\rm pre}}^2$, but not on resonator size. The Q-factors obtained in this Letter are comparable in magnitude to experimentally observed Q-factors \cite{eimo+11}. Therefore, it should be possible to experimentally verify the proposed mechanisms.

Since the system is closed, the rates reported here constitute a lower bound on the dissipation of graphene resonators. Many applications of nanomechanical systems rely on high Q-factors \cite{ilya+04,labu+04}, and therefore an improved understanding of the physical processes limiting this is of fundamental interest. Our results demonstrate the possibility of using graphene resonators as a test bed for FPU physics. By this approach long standing questions in non-equilibrium statistical mechanics might eventually be within experimental reach.

The authors acknowledge funding from the European Union through FP7 project no. 246026 (RODIN) (DM, AI), the Swedish Research Council (DM, AC), the Knut \& Alice Wallenberg foundation (DM, AI) and NSF CMMI-1036460 (ZQ,HSP). The authors acknowledge Prof. David Campbell for valuable input.

\bibliography{nldmgrex,bibFPU,bibmd}

%merlin.mbs apsrev4-1.bst 2010-07-25 4.21a (PWD, AO, DPC) hacked
%Control: key (0)
%Control: author (8) initials jnrlst
%Control: editor formatted (1) identically to author
%Control: production of article title (-1) disabled
%Control: page (0) single
%Control: year (1) truncated
%Control: production of eprint (0) enabled
\begin{thebibliography}{41}%
\makeatletter
\providecommand \@ifxundefined [1]{%
 \@ifx{#1\undefined}
}%
\providecommand \@ifnum [1]{%
 \ifnum #1\expandafter \@firstoftwo
 \else \expandafter \@secondoftwo
 \fi
}%
\providecommand \@ifx [1]{%
 \ifx #1\expandafter \@firstoftwo
 \else \expandafter \@secondoftwo
 \fi
}%
\providecommand \natexlab [1]{#1}%
\providecommand \enquote  [1]{``#1''}%
\providecommand \bibnamefont  [1]{#1}%
\providecommand \bibfnamefont [1]{#1}%
\providecommand \citenamefont [1]{#1}%
\providecommand \href@noop [0]{\@secondoftwo}%
\providecommand \href [0]{\begingroup \@sanitize@url \@href}%
\providecommand \@href[1]{\@@startlink{#1}\@@href}%
\providecommand \@@href[1]{\endgroup#1\@@endlink}%
\providecommand \@sanitize@url [0]{\catcode `\\12\catcode `\$12\catcode
  `\&12\catcode `\#12\catcode `\^12\catcode `\_12\catcode `\%12\relax}%
\providecommand \@@startlink[1]{}%
\providecommand \@@endlink[0]{}%
\providecommand \url  [0]{\begingroup\@sanitize@url \@url }%
\providecommand \@url [1]{\endgroup\@href {#1}{\urlprefix }}%
\providecommand \urlprefix  [0]{URL }%
\providecommand \Eprint [0]{\href }%
\providecommand \doibase [0]{http://dx.doi.org/}%
\providecommand \selectlanguage [0]{\@gobble}%
\providecommand \bibinfo  [0]{\@secondoftwo}%
\providecommand \bibfield  [0]{\@secondoftwo}%
\providecommand \translation [1]{[#1]}%
\providecommand \BibitemOpen [0]{}%
\providecommand \bibitemStop [0]{}%
\providecommand \bibitemNoStop [0]{.\EOS\space}%
\providecommand \EOS [0]{\spacefactor3000\relax}%
\providecommand \BibitemShut  [1]{\csname bibitem#1\endcsname}%
\let\auto@bib@innerbib\@empty
%</preamble>
\bibitem [{\citenamefont {Bunch}\ \emph {et~al.}(2007)\citenamefont {Bunch},
  \citenamefont {van~der Zande}, \citenamefont {Verbridge}, \citenamefont
  {Frank}, \citenamefont {Tanenbaum}, \citenamefont {Parpia}, \citenamefont
  {Craighead},\ and\ \citenamefont {McEuen}}]{buza+07}%
  \BibitemOpen
  \bibfield  {author} {\bibinfo {author} {\bibfnamefont {J.~S.}\ \bibnamefont
  {Bunch}}, \bibinfo {author} {\bibfnamefont {A.~M.}\ \bibnamefont {van~der
  Zande}}, \bibinfo {author} {\bibfnamefont {S.~S.}\ \bibnamefont {Verbridge}},
  \bibinfo {author} {\bibfnamefont {I.~W.}\ \bibnamefont {Frank}}, \bibinfo
  {author} {\bibfnamefont {D.~M.}\ \bibnamefont {Tanenbaum}}, \bibinfo {author}
  {\bibfnamefont {J.~M.}\ \bibnamefont {Parpia}}, \bibinfo {author}
  {\bibfnamefont {H.~G.}\ \bibnamefont {Craighead}}, \ and\ \bibinfo {author}
  {\bibfnamefont {P.~L.}\ \bibnamefont {McEuen}},\ }\href {\doibase
  10.1126/science.1136836} {\bibfield  {journal} {\bibinfo  {journal}
  {Science}\ }\textbf {\bibinfo {volume} {315}},\ \bibinfo {pages} {490}
  (\bibinfo {year} {2007})}\BibitemShut {NoStop}%
\bibitem [{\citenamefont {Eriksson}\ \emph {et~al.}(2008)\citenamefont
  {Eriksson}, \citenamefont {Lee}, \citenamefont {Sourab}, \citenamefont
  {Isacsson}, \citenamefont {Kaunisto}, \citenamefont {Kinaret},\ and\
  \citenamefont {Campbell}}]{erle+08}%
  \BibitemOpen
  \bibfield  {author} {\bibinfo {author} {\bibfnamefont {A.}~\bibnamefont
  {Eriksson}}, \bibinfo {author} {\bibfnamefont {S.}~\bibnamefont {Lee}},
  \bibinfo {author} {\bibfnamefont {A.~A.}\ \bibnamefont {Sourab}}, \bibinfo
  {author} {\bibfnamefont {A.}~\bibnamefont {Isacsson}}, \bibinfo {author}
  {\bibfnamefont {R.}~\bibnamefont {Kaunisto}}, \bibinfo {author}
  {\bibfnamefont {J.~M.}\ \bibnamefont {Kinaret}}, \ and\ \bibinfo {author}
  {\bibfnamefont {E.~E.~B.}\ \bibnamefont {Campbell}},\ }\href {\doibase
  10.1021/nl080345w} {\bibfield  {journal} {\bibinfo  {journal} {Nano Lett.}\
  }\textbf {\bibinfo {volume} {8}},\ \bibinfo {pages} {1224} (\bibinfo {year}
  {2008})}\BibitemShut {NoStop}%
\bibitem [{\citenamefont {Chen}\ \emph {et~al.}(2009)\citenamefont {Chen},
  \citenamefont {Rosenblatt}, \citenamefont {Bolotin}, \citenamefont {Kalb},
  \citenamefont {Kim}, \citenamefont {Kymissis}, \citenamefont {Stormer},
  \citenamefont {Heinz},\ and\ \citenamefont {Hone}}]{chro+09}%
  \BibitemOpen
  \bibfield  {author} {\bibinfo {author} {\bibfnamefont {C.}~\bibnamefont
  {Chen}}, \bibinfo {author} {\bibfnamefont {S.}~\bibnamefont {Rosenblatt}},
  \bibinfo {author} {\bibfnamefont {K.~I.}\ \bibnamefont {Bolotin}}, \bibinfo
  {author} {\bibfnamefont {W.}~\bibnamefont {Kalb}}, \bibinfo {author}
  {\bibfnamefont {P.}~\bibnamefont {Kim}}, \bibinfo {author} {\bibfnamefont
  {I.}~\bibnamefont {Kymissis}}, \bibinfo {author} {\bibfnamefont {H.~L.}\
  \bibnamefont {Stormer}}, \bibinfo {author} {\bibfnamefont {T.~F.}\
  \bibnamefont {Heinz}}, \ and\ \bibinfo {author} {\bibfnamefont
  {J.}~\bibnamefont {Hone}},\ }\href@noop {} {\bibfield  {journal} {\bibinfo
  {journal} {Nat. Nanotechnol.}\ }\textbf {\bibinfo {volume} {4}},\ \bibinfo
  {pages} {861} (\bibinfo {year} {2009})}\BibitemShut {NoStop}%
\bibitem [{\citenamefont {Eichler}\ \emph {et~al.}(2011)\citenamefont
  {Eichler}, \citenamefont {Moser}, \citenamefont {Chaste}, \citenamefont
  {Zdrojek}, \citenamefont {Wilson-Rae},\ and\ \citenamefont
  {Bachtold}}]{eimo+11}%
  \BibitemOpen
  \bibfield  {author} {\bibinfo {author} {\bibfnamefont {A.}~\bibnamefont
  {Eichler}}, \bibinfo {author} {\bibfnamefont {J.}~\bibnamefont {Moser}},
  \bibinfo {author} {\bibfnamefont {J.}~\bibnamefont {Chaste}}, \bibinfo
  {author} {\bibfnamefont {M.}~\bibnamefont {Zdrojek}}, \bibinfo {author}
  {\bibfnamefont {I.}~\bibnamefont {Wilson-Rae}}, \ and\ \bibinfo {author}
  {\bibfnamefont {A.}~\bibnamefont {Bachtold}},\ }\href {\doibase
  10.1038/nnano.2011.71} {\bibfield  {journal} {\bibinfo  {journal} {Nat.
  Nanotechnol.}\ }\textbf {\bibinfo {volume} {6}},\ \bibinfo {pages} {339}
  (\bibinfo {year} {2011})}\BibitemShut {NoStop}%
\bibitem [{\citenamefont {Oshidari}\ \emph {et~al.}(2012)\citenamefont
  {Oshidari}, \citenamefont {Hatakeyama}, \citenamefont {Kometani},
  \citenamefont {Warisawa},\ and\ \citenamefont {Ishihara}}]{osha+12}%
  \BibitemOpen
  \bibfield  {author} {\bibinfo {author} {\bibfnamefont {Y.}~\bibnamefont
  {Oshidari}}, \bibinfo {author} {\bibfnamefont {T.}~\bibnamefont
  {Hatakeyama}}, \bibinfo {author} {\bibfnamefont {R.}~\bibnamefont
  {Kometani}}, \bibinfo {author} {\bibfnamefont {S.}~\bibnamefont {Warisawa}},
  \ and\ \bibinfo {author} {\bibfnamefont {S.}~\bibnamefont {Ishihara}},\
  }\href@noop {} {\bibfield  {journal} {\bibinfo  {journal} {Appl. Phys.
  Express}\ }\textbf {\bibinfo {volume} {5}},\ \bibinfo {pages} {117201}
  (\bibinfo {year} {2012})}\BibitemShut {NoStop}%
\bibitem [{\citenamefont {Barton}\ \emph {et~al.}(2011)\citenamefont {Barton},
  \citenamefont {Ilic}, \citenamefont {van~der Zande}, \citenamefont {Whitney},
  \citenamefont {McEuen}, \citenamefont {Parpia},\ and\ \citenamefont
  {Craighead}}]{bail+11}%
  \BibitemOpen
  \bibfield  {author} {\bibinfo {author} {\bibfnamefont {R.~A.}\ \bibnamefont
  {Barton}}, \bibinfo {author} {\bibfnamefont {B.}~\bibnamefont {Ilic}},
  \bibinfo {author} {\bibfnamefont {A.~M.}\ \bibnamefont {van~der Zande}},
  \bibinfo {author} {\bibfnamefont {W.~S.}\ \bibnamefont {Whitney}}, \bibinfo
  {author} {\bibfnamefont {P.~L.}\ \bibnamefont {McEuen}}, \bibinfo {author}
  {\bibfnamefont {J.~M.}\ \bibnamefont {Parpia}}, \ and\ \bibinfo {author}
  {\bibfnamefont {H.~G.}\ \bibnamefont {Craighead}},\ }\href {\doibase
  10.1021/nl1042227} {\bibfield  {journal} {\bibinfo  {journal} {Nano Letters}\
  }\textbf {\bibinfo {volume} {11}},\ \bibinfo {pages} {1232} (\bibinfo {year}
  {2011})}\BibitemShut {NoStop}%
\bibitem [{\citenamefont {Zande}\ \emph {et~al.}(2010)\citenamefont {Zande},
  \citenamefont {Barton}, \citenamefont {Alden}, \citenamefont {Ruiz-Vargas},
  \citenamefont {Whitney}, \citenamefont {Pham}, \citenamefont {Park},
  \citenamefont {Parpia}, \citenamefont {Craighead},\ and\ \citenamefont
  {McEuen}}]{zaba+10}%
  \BibitemOpen
  \bibfield  {author} {\bibinfo {author} {\bibfnamefont {A.~M. v.~d.}\
  \bibnamefont {Zande}}, \bibinfo {author} {\bibfnamefont {R.~A.}\ \bibnamefont
  {Barton}}, \bibinfo {author} {\bibfnamefont {J.~S.}\ \bibnamefont {Alden}},
  \bibinfo {author} {\bibfnamefont {C.~S.}\ \bibnamefont {Ruiz-Vargas}},
  \bibinfo {author} {\bibfnamefont {W.~S.}\ \bibnamefont {Whitney}}, \bibinfo
  {author} {\bibfnamefont {P.~H.~Q.}\ \bibnamefont {Pham}}, \bibinfo {author}
  {\bibfnamefont {J.}~\bibnamefont {Park}}, \bibinfo {author} {\bibfnamefont
  {J.~M.}\ \bibnamefont {Parpia}}, \bibinfo {author} {\bibfnamefont {H.~G.}\
  \bibnamefont {Craighead}}, \ and\ \bibinfo {author} {\bibfnamefont {P.~L.}\
  \bibnamefont {McEuen}},\ }\href {\doibase 10.1021/nl102713c} {\bibfield
  {journal} {\bibinfo  {journal} {Nano Letters}\ }\textbf {\bibinfo {volume}
  {10}},\ \bibinfo {pages} {4869} (\bibinfo {year} {2010})}\BibitemShut
  {NoStop}%
\bibitem [{\citenamefont {Eom}\ \emph {et~al.}(2011)\citenamefont {Eom},
  \citenamefont {Park}, \citenamefont {Yoon},\ and\ \citenamefont
  {Kwon}}]{eopa+11}%
  \BibitemOpen
  \bibfield  {author} {\bibinfo {author} {\bibfnamefont {K.}~\bibnamefont
  {Eom}}, \bibinfo {author} {\bibfnamefont {H.~S.}\ \bibnamefont {Park}},
  \bibinfo {author} {\bibfnamefont {D.~S.}\ \bibnamefont {Yoon}}, \ and\
  \bibinfo {author} {\bibfnamefont {T.}~\bibnamefont {Kwon}},\ }\href {\doibase
  http://dx.doi.org/10.1016/j.physrep.2011.03.002} {\bibfield  {journal}
  {\bibinfo  {journal} {Physics Reports}\ }\textbf {\bibinfo {volume} {503}},\
  \bibinfo {pages} {115 } (\bibinfo {year} {2011})}\BibitemShut {NoStop}%
\bibitem [{\citenamefont {Atalaya}\ \emph {et~al.}(2008)\citenamefont
  {Atalaya}, \citenamefont {Isacsson},\ and\ \citenamefont
  {Kinaret}}]{atis+08}%
  \BibitemOpen
  \bibfield  {author} {\bibinfo {author} {\bibfnamefont {J.}~\bibnamefont
  {Atalaya}}, \bibinfo {author} {\bibfnamefont {A.}~\bibnamefont {Isacsson}}, \
  and\ \bibinfo {author} {\bibfnamefont {J.~M.}\ \bibnamefont {Kinaret}},\
  }\href {\doibase 10.1021/nl801733d} {\bibfield  {journal} {\bibinfo
  {journal} {Nano Lett.}\ }\textbf {\bibinfo {volume} {8}},\ \bibinfo {pages}
  {4196} (\bibinfo {year} {2008})}\BibitemShut {NoStop}%
\bibitem [{\citenamefont {Matheny}\ \emph {et~al.}(2013)\citenamefont
  {Matheny}, \citenamefont {Villanueva}, \citenamefont {Karabalin},
  \citenamefont {Sader},\ and\ \citenamefont {Roukes}}]{mavi+13}%
  \BibitemOpen
  \bibfield  {author} {\bibinfo {author} {\bibfnamefont {M.~H.}\ \bibnamefont
  {Matheny}}, \bibinfo {author} {\bibfnamefont {L.~G.}\ \bibnamefont
  {Villanueva}}, \bibinfo {author} {\bibfnamefont {R.~B.}\ \bibnamefont
  {Karabalin}}, \bibinfo {author} {\bibfnamefont {J.~E.}\ \bibnamefont
  {Sader}}, \ and\ \bibinfo {author} {\bibfnamefont {M.~L.}\ \bibnamefont
  {Roukes}},\ }\href {\doibase 10.1021/nl400070e} {\bibfield  {journal}
  {\bibinfo  {journal} {Nano Letters}\ }\textbf {\bibinfo {volume} {13}},\
  \bibinfo {pages} {1622} (\bibinfo {year} {2013})}\BibitemShut {NoStop}%
\bibitem [{\citenamefont {Eriksson}\ \emph {et~al.}(2013)\citenamefont
  {Eriksson}, \citenamefont {Midtvedt}, \citenamefont {Croy},\ and\
  \citenamefont {Isacsson}}]{ermi+13}%
  \BibitemOpen
  \bibfield  {author} {\bibinfo {author} {\bibfnamefont {A.~M.}\ \bibnamefont
  {Eriksson}}, \bibinfo {author} {\bibfnamefont {D.}~\bibnamefont {Midtvedt}},
  \bibinfo {author} {\bibfnamefont {A.}~\bibnamefont {Croy}}, \ and\ \bibinfo
  {author} {\bibfnamefont {A.}~\bibnamefont {Isacsson}},\ }\href@noop {}
  {\bibfield  {journal} {\bibinfo  {journal} {Nanotechnology}\ }\textbf
  {\bibinfo {volume} {24}},\ \bibinfo {pages} {39570} (\bibinfo {year}
  {2013})}\BibitemShut {NoStop}%
\bibitem [{\citenamefont {Lifshitz}\ and\ \citenamefont
  {Roukes}(2000)}]{liro00}%
  \BibitemOpen
  \bibfield  {author} {\bibinfo {author} {\bibfnamefont {R.}~\bibnamefont
  {Lifshitz}}\ and\ \bibinfo {author} {\bibfnamefont {M.~L.}\ \bibnamefont
  {Roukes}},\ }\href {\doibase 10.1103/PhysRevB.61.5600} {\bibfield  {journal}
  {\bibinfo  {journal} {Phys. Rev. B}\ }\textbf {\bibinfo {volume} {61}},\
  \bibinfo {pages} {5600} (\bibinfo {year} {2000})}\BibitemShut {NoStop}%
\bibitem [{\citenamefont {Cross}\ and\ \citenamefont
  {Lifshitz}(2001)}]{crli01}%
  \BibitemOpen
  \bibfield  {author} {\bibinfo {author} {\bibfnamefont {M.~C.}\ \bibnamefont
  {Cross}}\ and\ \bibinfo {author} {\bibfnamefont {R.}~\bibnamefont
  {Lifshitz}},\ }\href {\doibase 10.1103/PhysRevB.64.085324} {\bibfield
  {journal} {\bibinfo  {journal} {Phys. Rev. B}\ }\textbf {\bibinfo {volume}
  {64}},\ \bibinfo {pages} {085324} (\bibinfo {year} {2001})}\BibitemShut
  {NoStop}%
\bibitem [{\citenamefont {Wilson-Rae}(2008)}]{wil08}%
  \BibitemOpen
  \bibfield  {author} {\bibinfo {author} {\bibfnamefont {I.}~\bibnamefont
  {Wilson-Rae}},\ }\href {\doibase 10.1103/PhysRevB.77.245418} {\bibfield
  {journal} {\bibinfo  {journal} {Phys. Rev. B}\ }\textbf {\bibinfo {volume}
  {77}},\ \bibinfo {pages} {245418} (\bibinfo {year} {2008})}\BibitemShut
  {NoStop}%
\bibitem [{\citenamefont {Remus}\ \emph {et~al.}(2009)\citenamefont {Remus},
  \citenamefont {Blencowe},\ and\ \citenamefont {Tanaka}}]{rebl+09}%
  \BibitemOpen
  \bibfield  {author} {\bibinfo {author} {\bibfnamefont {L.~G.}\ \bibnamefont
  {Remus}}, \bibinfo {author} {\bibfnamefont {M.~P.}\ \bibnamefont {Blencowe}},
  \ and\ \bibinfo {author} {\bibfnamefont {Y.}~\bibnamefont {Tanaka}},\ }\href
  {\doibase 10.1103/PhysRevB.80.174103} {\bibfield  {journal} {\bibinfo
  {journal} {Phys. Rev. B}\ }\textbf {\bibinfo {volume} {80}},\ \bibinfo
  {pages} {174103} (\bibinfo {year} {2009})}\BibitemShut {NoStop}%
\bibitem [{\citenamefont {Croy}\ \emph {et~al.}(2012)\citenamefont {Croy},
  \citenamefont {Midtvedt}, \citenamefont {Isacsson},\ and\ \citenamefont
  {Kinaret}}]{crmi+12}%
  \BibitemOpen
  \bibfield  {author} {\bibinfo {author} {\bibfnamefont {A.}~\bibnamefont
  {Croy}}, \bibinfo {author} {\bibfnamefont {D.}~\bibnamefont {Midtvedt}},
  \bibinfo {author} {\bibfnamefont {A.}~\bibnamefont {Isacsson}}, \ and\
  \bibinfo {author} {\bibfnamefont {J.~M.}\ \bibnamefont {Kinaret}},\ }\href
  {\doibase 10.1103/PhysRevB.86.235435} {\bibfield  {journal} {\bibinfo
  {journal} {Phys. Rev. B}\ }\textbf {\bibinfo {volume} {86}},\ \bibinfo
  {pages} {235435} (\bibinfo {year} {2012})}\BibitemShut {NoStop}%
\bibitem [{\citenamefont {Imboden}\ and\ \citenamefont
  {Mohanty}(ress)}]{immo+13}%
  \BibitemOpen
  \bibfield  {author} {\bibinfo {author} {\bibfnamefont {M.}~\bibnamefont
  {Imboden}}\ and\ \bibinfo {author} {\bibfnamefont {P.}~\bibnamefont
  {Mohanty}},\ }\href {\doibase
  http://dx.doi.org/10.1016/j.physrep.2013.09.003} {\bibfield  {journal}
  {\bibinfo  {journal} {Physics Reports}\ ,\ } (\bibinfo {year} {in
  press})}\BibitemShut {NoStop}%
\bibitem [{\citenamefont {Barnard}\ \emph {et~al.}(2012)\citenamefont
  {Barnard}, \citenamefont {Sazonova}, \citenamefont {van~der Zande},\ and\
  \citenamefont {McEuen}}]{basa+12}%
  \BibitemOpen
  \bibfield  {author} {\bibinfo {author} {\bibfnamefont {A.~W.}\ \bibnamefont
  {Barnard}}, \bibinfo {author} {\bibfnamefont {V.}~\bibnamefont {Sazonova}},
  \bibinfo {author} {\bibfnamefont {A.~M.}\ \bibnamefont {van~der Zande}}, \
  and\ \bibinfo {author} {\bibfnamefont {P.~L.}\ \bibnamefont {McEuen}},\
  }\href {\doibase 10.1073/pnas.1216407109} {\bibfield  {journal} {\bibinfo
  {journal} {Proceedings of the National Academy of Sciences}\ }\textbf
  {\bibinfo {volume} {109}},\ \bibinfo {pages} {19093} (\bibinfo {year}
  {2012})}\BibitemShut {NoStop}%
\bibitem [{\citenamefont {Fermi}\ \emph {et~al.}(1955)\citenamefont {Fermi},
  \citenamefont {Pasta},\ and\ \citenamefont {Ulam}}]{FPU+55}%
  \BibitemOpen
  \bibfield  {author} {\bibinfo {author} {\bibfnamefont {E.}~\bibnamefont
  {Fermi}}, \bibinfo {author} {\bibfnamefont {J.~R.}\ \bibnamefont {Pasta}}, \
  and\ \bibinfo {author} {\bibfnamefont {S.}~\bibnamefont {Ulam}},\ }\href@noop
  {} {\emph {\bibinfo {title} {Studies of nonlinear problems}}},\ \bibinfo
  {type} {Tech. Rep.}\ (\bibinfo  {institution} {Los Alamos Scientific
  Laboratory},\ \bibinfo {year} {1955})\BibitemShut {NoStop}%
\bibitem [{\citenamefont {Campbell}\ \emph {et~al.}(2005)\citenamefont
  {Campbell}, \citenamefont {Rosenau},\ and\ \citenamefont
  {Zaslavsky}}]{ca+05}%
  \BibitemOpen
  \bibfield  {author} {\bibinfo {author} {\bibfnamefont {D.~K.}\ \bibnamefont
  {Campbell}}, \bibinfo {author} {\bibfnamefont {P.}~\bibnamefont {Rosenau}}, \
  and\ \bibinfo {author} {\bibfnamefont {G.}~\bibnamefont {Zaslavsky}},\
  }\href@noop {} {\bibfield  {journal} {\bibinfo  {journal} {Chaos}\ }
  (\bibinfo {year} {2005})}\BibitemShut {NoStop}%
\bibitem [{\citenamefont {Izrailev}\ and\ \citenamefont
  {Chirikov}(1966)}]{izch+66}%
  \BibitemOpen
  \bibfield  {author} {\bibinfo {author} {\bibfnamefont {F.}~\bibnamefont
  {Izrailev}}\ and\ \bibinfo {author} {\bibfnamefont {B.~V.}\ \bibnamefont
  {Chirikov}},\ }\href@noop {} {\bibfield  {journal} {\bibinfo  {journal}
  {Soviet Physics Doklady}\ }\textbf {\bibinfo {volume} {11}},\ \bibinfo
  {pages} {30} (\bibinfo {year} {1966})}\BibitemShut {NoStop}%
\bibitem [{\citenamefont {Zabusky}\ and\ \citenamefont
  {Kruskal}(1965)}]{zakr+65}%
  \BibitemOpen
  \bibfield  {author} {\bibinfo {author} {\bibfnamefont {N.}~\bibnamefont
  {Zabusky}}\ and\ \bibinfo {author} {\bibfnamefont {M.~D.}\ \bibnamefont
  {Kruskal}},\ }\href@noop {} {\bibfield  {journal} {\bibinfo  {journal}
  {Physical Review Letters}\ }\textbf {\bibinfo {volume} {15}},\ \bibinfo
  {pages} {240} (\bibinfo {year} {1965})}\BibitemShut {NoStop}%
\bibitem [{\citenamefont {Penati}\ and\ \citenamefont {Flach}(2007)}]{pefl+07}%
  \BibitemOpen
  \bibfield  {author} {\bibinfo {author} {\bibfnamefont {T.}~\bibnamefont
  {Penati}}\ and\ \bibinfo {author} {\bibfnamefont {S.}~\bibnamefont {Flach}},\
  }\href@noop {} {\bibfield  {journal} {\bibinfo  {journal} {Chaos}\ }\textbf
  {\bibinfo {volume} {17}} (\bibinfo {year} {2007})}\BibitemShut {NoStop}%
\bibitem [{\citenamefont {Flach}\ \emph {et~al.}(2005)\citenamefont {Flach},
  \citenamefont {Ivanchenko},\ and\ \citenamefont {Kanakov}}]{fliv+05}%
  \BibitemOpen
  \bibfield  {author} {\bibinfo {author} {\bibfnamefont {S.}~\bibnamefont
  {Flach}}, \bibinfo {author} {\bibfnamefont {M.~V.}\ \bibnamefont
  {Ivanchenko}}, \ and\ \bibinfo {author} {\bibfnamefont {O.~I.}\ \bibnamefont
  {Kanakov}},\ }\href@noop {} {\bibfield  {journal} {\bibinfo  {journal} {Phys.
  Rev. Lett.}\ }\textbf {\bibinfo {volume} {95}},\ \bibinfo {pages} {064102}
  (\bibinfo {year} {2005})}\BibitemShut {NoStop}%
\bibitem [{\citenamefont {Fucito}\ \emph {et~al.}(1982)\citenamefont {Fucito},
  \citenamefont {Marchesoni}, \citenamefont {Marinari}, \citenamefont {Parisi},
  \citenamefont {Peliti},\ and\ \citenamefont {Ruffo}}]{fuma+82}%
  \BibitemOpen
  \bibfield  {author} {\bibinfo {author} {\bibfnamefont {F.}~\bibnamefont
  {Fucito}}, \bibinfo {author} {\bibnamefont {Marchesoni}}, \bibinfo {author}
  {\bibfnamefont {F.}~\bibnamefont {Marinari}}, \bibinfo {author}
  {\bibfnamefont {G.}~\bibnamefont {Parisi}}, \bibinfo {author} {\bibfnamefont
  {L.}~\bibnamefont {Peliti}}, \ and\ \bibinfo {author} {\bibfnamefont
  {S.}~\bibnamefont {Ruffo}},\ }\href@noop {} {\bibfield  {journal} {\bibinfo
  {journal} {Le Journal de Physique}\ } (\bibinfo {year} {1982})}\BibitemShut
  {NoStop}%
\bibitem [{\citenamefont {Bambusi}\ and\ \citenamefont
  {Ponno}(2008)}]{bapo+08}%
  \BibitemOpen
  \bibfield  {author} {\bibinfo {author} {\bibfnamefont {D.}~\bibnamefont
  {Bambusi}}\ and\ \bibinfo {author} {\bibfnamefont {A.}~\bibnamefont
  {Ponno}},\ }in\ \href@noop {} {\emph {\bibinfo {booktitle} {The
  Fermi-Pasta-Ulam problem}}}\ (\bibinfo  {publisher} {Springer},\ \bibinfo
  {year} {2008})\BibitemShut {NoStop}%
\bibitem [{\citenamefont {Benettin}\ \emph {et~al.}(2008)\citenamefont
  {Benettin}, \citenamefont {Carati}, \citenamefont {Galgani},\ and\
  \citenamefont {Giorgilli}}]{beca+08}%
  \BibitemOpen
  \bibfield  {author} {\bibinfo {author} {\bibfnamefont {G.}~\bibnamefont
  {Benettin}}, \bibinfo {author} {\bibfnamefont {A.}~\bibnamefont {Carati}},
  \bibinfo {author} {\bibfnamefont {L.}~\bibnamefont {Galgani}}, \ and\
  \bibinfo {author} {\bibfnamefont {A.}~\bibnamefont {Giorgilli}},\ }in\
  \href@noop {} {\emph {\bibinfo {booktitle} {The Fermi-Pasta-Ulam problem}}}\
  (\bibinfo  {publisher} {Springer},\ \bibinfo {year} {2008})\BibitemShut
  {NoStop}%
\bibitem [{\citenamefont {Lammps}(2012)}]{plimptonLAMMPS}%
  \BibitemOpen
  \bibfield  {author} {\bibinfo {author} {\bibnamefont {Lammps}},\ }\href@noop
  {} {\bibfield  {journal} {\bibinfo  {journal} {http://lammps.sandia.gov}\ }
  (\bibinfo {year} {2012})}\BibitemShut {NoStop}%
\bibitem [{\citenamefont {Plimpton}(1995)}]{plimptonJCP1995}%
  \BibitemOpen
  \bibfield  {author} {\bibinfo {author} {\bibfnamefont {S.~J.}\ \bibnamefont
  {Plimpton}},\ }\href@noop {} {\bibfield  {journal} {\bibinfo  {journal}
  {Journal of Computational Physics}\ }\textbf {\bibinfo {volume} {117}},\
  \bibinfo {pages} {1} (\bibinfo {year} {1995})}\BibitemShut {NoStop}%
\bibitem [{\citenamefont {Stuart}\ \emph {et~al.}(2000)\citenamefont {Stuart},
  \citenamefont {Tutein},\ and\ \citenamefont {Harrison}}]{stuartJCP2000}%
  \BibitemOpen
  \bibfield  {author} {\bibinfo {author} {\bibfnamefont {S.~J.}\ \bibnamefont
  {Stuart}}, \bibinfo {author} {\bibfnamefont {A.~B.}\ \bibnamefont {Tutein}},
  \ and\ \bibinfo {author} {\bibfnamefont {J.~A.}\ \bibnamefont {Harrison}},\
  }\href@noop {} {\bibfield  {journal} {\bibinfo  {journal} {Journal of
  Chemical Physics}\ }\textbf {\bibinfo {volume} {112}},\ \bibinfo {pages}
  {6472} (\bibinfo {year} {2000})}\BibitemShut {NoStop}%
\bibitem [{\citenamefont {Zhao}\ \emph {et~al.}(2009)\citenamefont {Zhao},
  \citenamefont {Min},\ and\ \citenamefont {Aluru}}]{zhaoNL2009}%
  \BibitemOpen
  \bibfield  {author} {\bibinfo {author} {\bibfnamefont {H.}~\bibnamefont
  {Zhao}}, \bibinfo {author} {\bibfnamefont {K.}~\bibnamefont {Min}}, \ and\
  \bibinfo {author} {\bibfnamefont {N.~R.}\ \bibnamefont {Aluru}},\ }\href@noop
  {} {\bibfield  {journal} {\bibinfo  {journal} {Nano Letters}\ }\textbf
  {\bibinfo {volume} {9}},\ \bibinfo {pages} {3012} (\bibinfo {year}
  {2009})}\BibitemShut {NoStop}%
\bibitem [{\citenamefont {Qi}\ and\ \citenamefont {Park}(2012)}]{qiNS2012}%
  \BibitemOpen
  \bibfield  {author} {\bibinfo {author} {\bibfnamefont {Z.}~\bibnamefont
  {Qi}}\ and\ \bibinfo {author} {\bibfnamefont {H.~S.}\ \bibnamefont {Park}},\
  }\href@noop {} {\bibfield  {journal} {\bibinfo  {journal} {Nanoscale}\
  }\textbf {\bibinfo {volume} {4}},\ \bibinfo {pages} {3460} (\bibinfo {year}
  {2012})}\BibitemShut {NoStop}%
\bibitem [{\citenamefont {Qi}\ \emph {et~al.}(2010)\citenamefont {Qi},
  \citenamefont {Zhao}, \citenamefont {Zhou}, \citenamefont {Sun},
  \citenamefont {Park},\ and\ \citenamefont {Wu}}]{qiNANO2010}%
  \BibitemOpen
  \bibfield  {author} {\bibinfo {author} {\bibfnamefont {Z.}~\bibnamefont
  {Qi}}, \bibinfo {author} {\bibfnamefont {F.}~\bibnamefont {Zhao}}, \bibinfo
  {author} {\bibfnamefont {X.}~\bibnamefont {Zhou}}, \bibinfo {author}
  {\bibfnamefont {Z.}~\bibnamefont {Sun}}, \bibinfo {author} {\bibfnamefont
  {H.~S.}\ \bibnamefont {Park}}, \ and\ \bibinfo {author} {\bibfnamefont
  {H.}~\bibnamefont {Wu}},\ }\href@noop {} {\bibfield  {journal} {\bibinfo
  {journal} {Nanotechnology}\ }\textbf {\bibinfo {volume} {21}},\ \bibinfo
  {pages} {265702} (\bibinfo {year} {2010})}\BibitemShut {NoStop}%
\bibitem [{\citenamefont {Lichtenberg}\ \emph {et~al.}(2008)\citenamefont
  {Lichtenberg}, \citenamefont {Livi},\ and\ \citenamefont
  {Pettini}}]{lili+08}%
  \BibitemOpen
  \bibfield  {author} {\bibinfo {author} {\bibfnamefont {A.~J.}\ \bibnamefont
  {Lichtenberg}}, \bibinfo {author} {\bibfnamefont {R.}~\bibnamefont {Livi}}, \
  and\ \bibinfo {author} {\bibfnamefont {M.}~\bibnamefont {Pettini}},\ }in\
  \href@noop {} {\emph {\bibinfo {booktitle} {The Fermi-Pasta-Ulam problem}}},\
  \bibinfo {editor} {edited by\ \bibinfo {editor} {\bibfnamefont
  {G.}~\bibnamefont {Gallavotti}}}\ (\bibinfo  {publisher} {Springer},\
  \bibinfo {year} {2008})\BibitemShut {NoStop}%
\bibitem [{\citenamefont {Lee}\ \emph {et~al.}(2008)\citenamefont {Lee},
  \citenamefont {Wei}, \citenamefont {Kysar},\ and\ \citenamefont
  {Hone}}]{lewe+08}%
  \BibitemOpen
  \bibfield  {author} {\bibinfo {author} {\bibfnamefont {C.}~\bibnamefont
  {Lee}}, \bibinfo {author} {\bibfnamefont {X.}~\bibnamefont {Wei}}, \bibinfo
  {author} {\bibfnamefont {J.~W.}\ \bibnamefont {Kysar}}, \ and\ \bibinfo
  {author} {\bibfnamefont {J.}~\bibnamefont {Hone}},\ }\href {\doibase
  10.1126/science.1157996} {\bibfield  {journal} {\bibinfo  {journal}
  {Science}\ }\textbf {\bibinfo {volume} {321}},\ \bibinfo {pages} {385}
  (\bibinfo {year} {2008})}\BibitemShut {NoStop}%
\bibitem [{\citenamefont {Benettin}\ \emph {et~al.}(1984)\citenamefont
  {Benettin}, \citenamefont {Galgani},\ and\ \citenamefont
  {Giorgilli}}]{bega+84}%
  \BibitemOpen
  \bibfield  {author} {\bibinfo {author} {\bibfnamefont {G.}~\bibnamefont
  {Benettin}}, \bibinfo {author} {\bibfnamefont {L.}~\bibnamefont {Galgani}}, \
  and\ \bibinfo {author} {\bibfnamefont {A.}~\bibnamefont {Giorgilli}},\ }\href
  {http://dx.doi.org/10.1038/311444a0} {\bibfield  {journal} {\bibinfo
  {journal} {Nature}\ }\textbf {\bibinfo {volume} {311}},\ \bibinfo {pages}
  {444} (\bibinfo {year} {1984})}\BibitemShut {NoStop}%
\bibitem [{\citenamefont {Pettini}\ and\ \citenamefont
  {Landolfi}(1990)}]{pela+89}%
  \BibitemOpen
  \bibfield  {author} {\bibinfo {author} {\bibfnamefont {M.}~\bibnamefont
  {Pettini}}\ and\ \bibinfo {author} {\bibfnamefont {M.}~\bibnamefont
  {Landolfi}},\ }\href@noop {} {\bibfield  {journal} {\bibinfo  {journal}
  {Phys. Rev. A}\ }\textbf {\bibinfo {volume} {41}},\ \bibinfo {pages} {768}
  (\bibinfo {year} {1990})}\BibitemShut {NoStop}%
\bibitem [{\citenamefont {Berchialla}\ \emph {et~al.}(2004)\citenamefont
  {Berchialla}, \citenamefont {Giorgilli},\ and\ \citenamefont
  {Paleari}}]{begi+04}%
  \BibitemOpen
  \bibfield  {author} {\bibinfo {author} {\bibfnamefont {L.}~\bibnamefont
  {Berchialla}}, \bibinfo {author} {\bibfnamefont {A.}~\bibnamefont
  {Giorgilli}}, \ and\ \bibinfo {author} {\bibfnamefont {S.}~\bibnamefont
  {Paleari}},\ }\href {\doibase
  http://dx.doi.org/10.1016/j.physleta.2003.11.052} {\bibfield  {journal}
  {\bibinfo  {journal} {Physics Letters A}\ }\textbf {\bibinfo {volume}
  {321}},\ \bibinfo {pages} {167 } (\bibinfo {year} {2004})}\BibitemShut
  {NoStop}%
\bibitem [{\citenamefont {Berman}\ and\ \citenamefont
  {Izrailev}(2005)}]{beiz+05}%
  \BibitemOpen
  \bibfield  {author} {\bibinfo {author} {\bibfnamefont {G.~P.}\ \bibnamefont
  {Berman}}\ and\ \bibinfo {author} {\bibfnamefont {F.~M.}\ \bibnamefont
  {Izrailev}},\ }\href@noop {} {\bibfield  {journal} {\bibinfo  {journal}
  {Chaos}\ }\textbf {\bibinfo {volume} {15}} (\bibinfo {year}
  {2005})}\BibitemShut {NoStop}%
\bibitem [{\citenamefont {Ilic}\ \emph {et~al.}(2004)\citenamefont {Ilic},
  \citenamefont {Yang},\ and\ \citenamefont {Craighead}}]{ilya+04}%
  \BibitemOpen
  \bibfield  {author} {\bibinfo {author} {\bibfnamefont {B.}~\bibnamefont
  {Ilic}}, \bibinfo {author} {\bibfnamefont {Y.}~\bibnamefont {Yang}}, \ and\
  \bibinfo {author} {\bibfnamefont {H.}~\bibnamefont {Craighead}},\ }\href
  {\doibase 10.1063/1.1794378} {\bibfield  {journal} {\bibinfo  {journal}
  {Applied Physics Letters}\ }\textbf {\bibinfo {volume} {85}},\ \bibinfo
  {pages} {2604} (\bibinfo {year} {2004})}\BibitemShut {NoStop}%
\bibitem [{\citenamefont {LaHaye}\ \emph {et~al.}(2004)\citenamefont {LaHaye},
  \citenamefont {Buu}, \citenamefont {Camarota},\ and\ \citenamefont
  {Schwab}}]{labu+04}%
  \BibitemOpen
  \bibfield  {author} {\bibinfo {author} {\bibfnamefont {M.~D.}\ \bibnamefont
  {LaHaye}}, \bibinfo {author} {\bibfnamefont {O.}~\bibnamefont {Buu}},
  \bibinfo {author} {\bibfnamefont {B.}~\bibnamefont {Camarota}}, \ and\
  \bibinfo {author} {\bibfnamefont {K.~C.}\ \bibnamefont {Schwab}},\ }\href
  {\doibase 10.1126/science.1094419} {\bibfield  {journal} {\bibinfo  {journal}
  {Science}\ }\textbf {\bibinfo {volume} {304}},\ \bibinfo {pages} {74}
  (\bibinfo {year} {2004})}\BibitemShut {NoStop}%
\end{thebibliography}%


%merlin.mbs apsrev4-1.bst 2010-07-25 4.21a (PWD, AO, DPC) hacked
%Control: key (0)
%Control: author (8) initials jnrlst
%Control: editor formatted (1) identically to author
%Control: production of article title (-1) disabled
%Control: page (0) single
%Control: year (1) truncated
%Control: production of eprint (0) enabled
\begin{thebibliography}{5}%
\makeatletter
\providecommand \@ifxundefined [1]{%
 \@ifx{#1\undefined}
}%
\providecommand \@ifnum [1]{%
 \ifnum #1\expandafter \@firstoftwo
 \else \expandafter \@secondoftwo
 \fi
}%
\providecommand \@ifx [1]{%
 \ifx #1\expandafter \@firstoftwo
 \else \expandafter \@secondoftwo
 \fi
}%
\providecommand \natexlab [1]{#1}%
\providecommand \enquote  [1]{``#1''}%
\providecommand \bibnamefont  [1]{#1}%
\providecommand \bibfnamefont [1]{#1}%
\providecommand \citenamefont [1]{#1}%
\providecommand \href@noop [0]{\@secondoftwo}%
\providecommand \href [0]{\begingroup \@sanitize@url \@href}%
\providecommand \@href[1]{\@@startlink{#1}\@@href}%
\providecommand \@@href[1]{\endgroup#1\@@endlink}%
\providecommand \@sanitize@url [0]{\catcode `\\12\catcode `\$12\catcode
  `\&12\catcode `\#12\catcode `\^12\catcode `\_12\catcode `\%12\relax}%
\providecommand \@@startlink[1]{}%
\providecommand \@@endlink[0]{}%
\providecommand \url  [0]{\begingroup\@sanitize@url \@url }%
\providecommand \@url [1]{\endgroup\@href {#1}{\urlprefix }}%
\providecommand \urlprefix  [0]{URL }%
\providecommand \Eprint [0]{\href }%
\providecommand \doibase [0]{http://dx.doi.org/}%
\providecommand \selectlanguage [0]{\@gobble}%
\providecommand \bibinfo  [0]{\@secondoftwo}%
\providecommand \bibfield  [0]{\@secondoftwo}%
\providecommand \translation [1]{[#1]}%
\providecommand \BibitemOpen [0]{}%
\providecommand \bibitemStop [0]{}%
\providecommand \bibitemNoStop [0]{.\EOS\space}%
\providecommand \EOS [0]{\spacefactor3000\relax}%
\providecommand \BibitemShut  [1]{\csname bibitem#1\endcsname}%
\let\auto@bib@innerbib\@empty
%</preamble>
\bibitem [{\citenamefont {Atalaya}\ \emph {et~al.}(2008)\citenamefont
  {Atalaya}, \citenamefont {Isacsson},\ and\ \citenamefont
  {Kinaret}}]{atis+08}%
  \BibitemOpen
  \bibfield  {author} {\bibinfo {author} {\bibfnamefont {J.}~\bibnamefont
  {Atalaya}}, \bibinfo {author} {\bibfnamefont {A.}~\bibnamefont {Isacsson}}, \
  and\ \bibinfo {author} {\bibfnamefont {J.~M.}\ \bibnamefont {Kinaret}},\
  }\href {\doibase 10.1021/nl801733d} {\bibfield  {journal} {\bibinfo
  {journal} {Nano Lett.}\ }\textbf {\bibinfo {volume} {8}},\ \bibinfo {pages}
  {4196} (\bibinfo {year} {2008})}\BibitemShut {NoStop}%
\bibitem [{\citenamefont {Eriksson}\ \emph {et~al.}(2013)\citenamefont
  {Eriksson}, \citenamefont {Midtvedt}, \citenamefont {Croy},\ and\
  \citenamefont {Isacsson}}]{ermi+13}%
  \BibitemOpen
  \bibfield  {author} {\bibinfo {author} {\bibfnamefont {A.~M.}\ \bibnamefont
  {Eriksson}}, \bibinfo {author} {\bibfnamefont {D.}~\bibnamefont {Midtvedt}},
  \bibinfo {author} {\bibfnamefont {A.}~\bibnamefont {Croy}}, \ and\ \bibinfo
  {author} {\bibfnamefont {A.}~\bibnamefont {Isacsson}},\ }\href@noop {}
  {\bibfield  {journal} {\bibinfo  {journal} {Nanotechnology}\ }\textbf
  {\bibinfo {volume} {24}},\ \bibinfo {pages} {39570} (\bibinfo {year}
  {2013})}\BibitemShut {NoStop}%
\bibitem [{\citenamefont {Zabusky}\ and\ \citenamefont
  {Kruskal}(1965)}]{zakr+65}%
  \BibitemOpen
  \bibfield  {author} {\bibinfo {author} {\bibfnamefont {N.}~\bibnamefont
  {Zabusky}}\ and\ \bibinfo {author} {\bibfnamefont {M.~D.}\ \bibnamefont
  {Kruskal}},\ }\href@noop {} {\bibfield  {journal} {\bibinfo  {journal}
  {Physical Review Letters}\ }\textbf {\bibinfo {volume} {15}},\ \bibinfo
  {pages} {240} (\bibinfo {year} {1965})}\BibitemShut {NoStop}%
\bibitem [{\citenamefont {Izrailev}\ and\ \citenamefont
  {Chirikov}(1966)}]{izch+66}%
  \BibitemOpen
  \bibfield  {author} {\bibinfo {author} {\bibfnamefont {F.}~\bibnamefont
  {Izrailev}}\ and\ \bibinfo {author} {\bibfnamefont {B.~V.}\ \bibnamefont
  {Chirikov}},\ }\href@noop {} {\bibfield  {journal} {\bibinfo  {journal}
  {Soviet Physics Doklady}\ }\textbf {\bibinfo {volume} {11}},\ \bibinfo
  {pages} {30} (\bibinfo {year} {1966})}\BibitemShut {NoStop}%
\bibitem [{\citenamefont {Flach}\ \emph {et~al.}(2005)\citenamefont {Flach},
  \citenamefont {Ivanchenko},\ and\ \citenamefont {Kanakov}}]{fliv+05}%
  \BibitemOpen
  \bibfield  {author} {\bibinfo {author} {\bibfnamefont {S.}~\bibnamefont
  {Flach}}, \bibinfo {author} {\bibfnamefont {M.~V.}\ \bibnamefont
  {Ivanchenko}}, \ and\ \bibinfo {author} {\bibfnamefont {O.~I.}\ \bibnamefont
  {Kanakov}},\ }\href@noop {} {\bibfield  {journal} {\bibinfo  {journal} {Phys.
  Rev. Lett.}\ }\textbf {\bibinfo {volume} {95}} (\bibinfo {year}
  {2005})}\BibitemShut {NoStop}%
\end{thebibliography}%
\end{document}

\title{Supplementary information for: FPU physics with nanomechanical graphene resonators: intrinsic relaxation and thermalization from flexural mode coupling}
	\author{Daniel Midtvedt}
	\email{midtvedt@chalmers.se}
	\affiliation{Department of Applied Physics, Chalmers University of Technology, 
					SE-412 96 G\"{o}teborg, Sweden}
	\author{Zenan Qi}
	\affiliation{Department of Mechanical Engineering, Boston University, Boston, MA 02215, USA}
	\author{Alexander Croy}
	\affiliation{Department of Applied Physics, Chalmers University of Technology, 
					SE-412 96 G\"{o}teborg, Sweden}
	\author{Harold S. Park}
	\affiliation{Department of Mechanical Engineering, Boston University, Boston, MA 02215, USA}
	\author{Andreas Isacsson}
	\affiliation{Department of Applied Physics, Chalmers University of Technology, 
					SE-412 96 G\"{o}teborg, Sweden}
\maketitle
\section*{Derivation of coupled mode equations}
\begin{figure}
		\centering
		\includegraphics[width=.7\columnwidth]{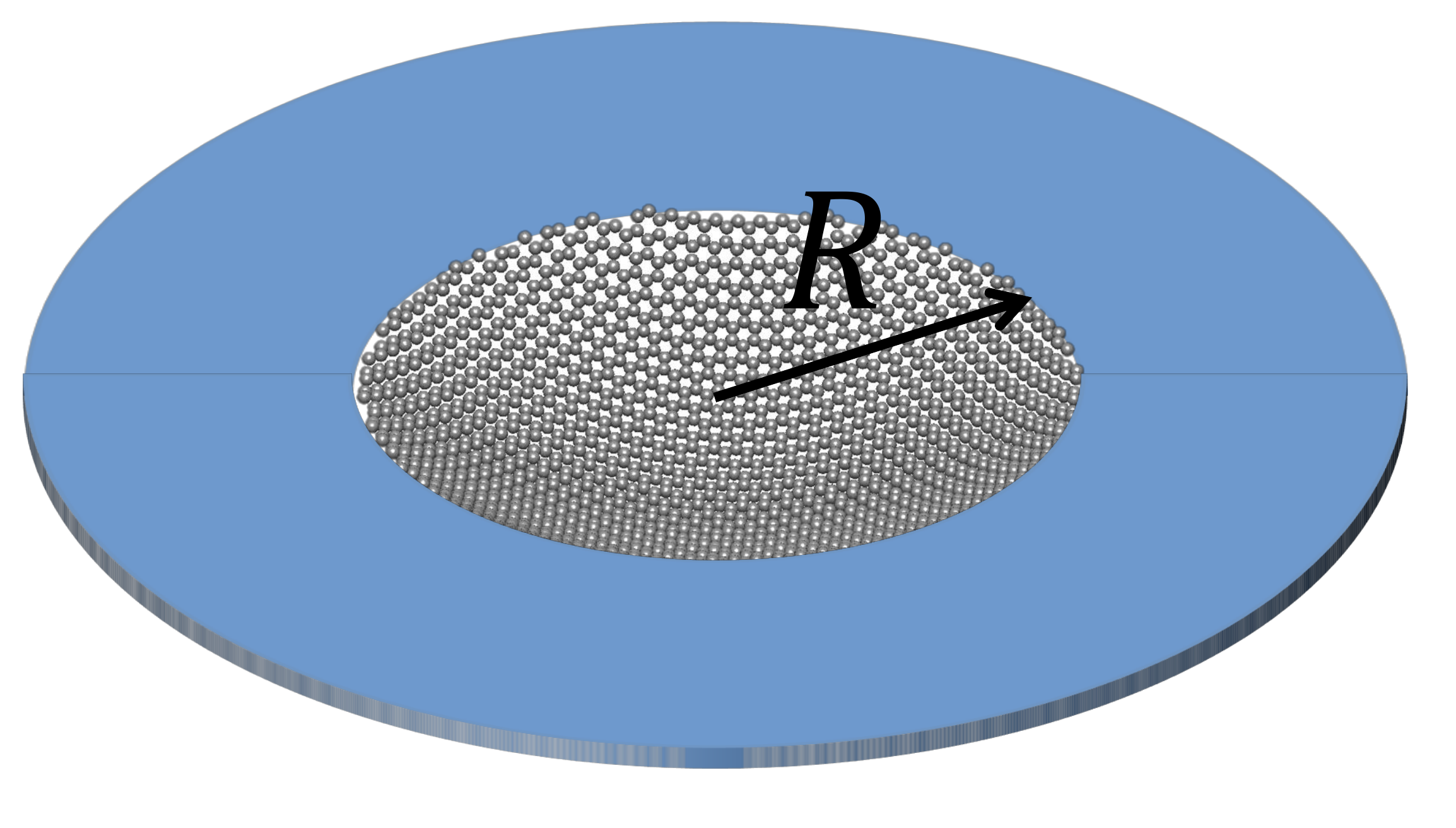}\hfill
				
\caption{Schematic image of the simulated structure.\label{prototype}}
	\end{figure}

The dynamics of suspended graphene resonators are described within continuum mechanics by the  F\"{o}ppl-von Karmann equations \cite{atis+08}. For the drum geometry (Fig. \ref{prototype}), they read \cite{ermi+13}
\begin{subequations}
\begin{equation}
\rhoG \ddot{u}_r-\left[\partial_r \sigma_{rr}+r^{-1}\partial_{\phi}
    \sigma_{r\phi}+r^{-1}(\sigma_{rr}-\sigma_{\phi\phi})
    \right]=0,\label{reqfirst}
    \end{equation} 
    \begin{equation} \rhoG
  \ddot{u}_{\phi}-\left[\partial_r
    \sigma_{r\phi}+{2}r^{-1}\sigma_{r\phi}+r^{-1}\partial_{\phi}\sigma_{\phi\phi}
    \right]=0,\label{phieqfirst}
    \end{equation}
    \begin{equation}
   \rhoG
  \ddot{w} - \epsilon_{\rm pre} (\lambda + 2\mu) r^{-1}\partial_r(r \partial_r w)  -  r^{-1}\left[\partial_r(r\sigma_{rr}
    \partial_r
    w+\sigma_{r\phi}\partial_{\phi}w) 
    +\partial_{\phi}\left(\sigma_{r\phi}\partial_r
    w +r^{-1}\sigma_{\phi\phi}\partial_{\phi}w\right)
    \right]=0,\label{weqfirst}
    \end{equation}
\end{subequations}
where $u_r$, $u_{\phi}$ and $w$ are the radial, angular and out-of-plane displacement fields, $\sigma_{ij}$ is the stress tensor, $\rhoG$ is the mass density of graphene, $\epsilon_{\rm pre}$ is the pre-strain of the sheet and $\lambda$ and $\mu$ are the Lam\'e parameters of graphene. To derive the equation of motion for the mode dynamics,  the in-plane dynamics is first eliminated adiabatically by setting $\ddot{u}_r=0$ and $\ddot{u}_{\phi}=0$ and solve the remaining stationary equations for the in-plane fields. The equations are made dimensionless through the following scaled variables, 
\begin{equation}
\tilde{r} = \frac{r}{R};\quad 
\tilde{t} = \frac{\sqrt{\epsilon_{\rm pre}} c_{\rm L}}{R} t; \quad 
\tilde{w}^2 = \frac{1}{2} \frac{w^2}{R^2 \epsilon_{\rm pre}};\quad
\tilde{E} = \frac{E}{2\epsilon_{\rm pre}^2 c_{\rm L}^2 \rho_{\rm G} R^2},%\quad
%\Phi_z =P_z{R}/{T_{\rm 0}},
\end{equation}
where $c_{\rm L}^2=(\lambda +2\mu)/\rho_{\rm G}$ is the longitudinal speed of sound in graphene and $R$ is the radius of the drum. In the following, only radially symmetric deformations are considered.

The dimensionless equations for the in- and out-of-plane components are, skipping the tildes over the scaled variables for convenience,
\begin{subequations}
\begin{equation}
\left[u_{r,rr}+\frac{1}{r} u_{r,r} -\frac{1}{r^2} u_r\right]-\frac{1-\nu}{2} \frac{1}{r} w_{,r}^2 - \frac{1}{2}\partial_r w_{,r}^2 =0,
\end{equation}
\begin{equation}
\ddot{w}-\left(\partial_r + \frac{1}{r}\right) \left(w_{,r} + \frac{w_{,r}^3}{2}\right)-\left(\partial_r + \frac{1}{r}\right)u_{r,r} w_{,r} -\nu \frac{1}{r}\partial_r u_r w_{,r} =0,
\end{equation}
\end{subequations}
with boundary conditions $w(1,t)=0$, $u_r(1,t)=0$. Subscript $,r$ denotes radial differentiation and $\nu$ is the Poisson ratio. The in-plane field is expanded as $u_r (r,t)=\sum_k Q_k(t) J_1(\xi_{1k} r)$ where $\xi_{nk}$ is the k:th zero of the n:th Bessel function, $J_n(\xi_{nk})=0$. The amplitudes $Q_k(t)$ are given by

\begin{equation}
Q_k= \frac{J_2(\xi_{1k})^2}{2 \xi_{1k}^2} C_k(t),
\end{equation}
where
\begin{equation}
C_k(t)\equiv\int_{0}^{1} dr r J_1(\xi_{1k} r) \left[\frac{1-\nu}{2r}w_{,r}^2 + \frac{1}{2}\partial_r w_{,r}^2\right].
\end{equation}

Inserting this result into the equation for $w$ we obtain
\begin{equation}
\ddot{w}-\left(\partial_r + \frac{1}{r}\right) \left(w_{,r} + \frac{w_{,r}^3}{2}\right)-
\sum_k \frac{J_2(\xi_{1k})^2C_k}{2\xi_{1k}^2}\left[\left(\partial_r+\frac{1}{r}\right)J'_1(\xi_{1k} r) w_{,r} + \frac{\nu}{r }\partial_r\left(J_1(\xi_{1k} r) w_{,r}\right)\right]=0.
\end{equation}
Next we expand $w(r,t)=\sum_n q_n(t) J_0(\xi_{0n} r)\equiv \sum_n q_n(t) \phi_n( r)$ for brevity. The equation of motion then becomes
\begin{equation}
 \frac{2}{J_1(\xi_{0n})^2}\ddot{q}_n +  \frac{2}{J_1(\xi_{0n})^2}  \xi_{0n}^2 q_n + \frac{1}{2}\sum_{k,l,m} M_{klmn} q_k q_l q_m - \sum_{k,l} \frac{J_2(\xi_{1k})^2C_k}{2\xi_{1k}^2} K_{kln} q_l,
\end{equation}
with
\begin{subequations}
\begin{equation}
M_{klmn}=\int_0^1 dr r \phi_{k,r} \phi_{l,r} \phi_{m,r} \phi_{n,r},
\end{equation}
\begin{equation}
K_{kln}=-\int_0^1 dr\left[ r J'_1(\xi_{1,k} r)\phi_{n,r}\phi_{l,r} + \nu J_1(\xi_{1,k} r)\phi_{n,r}\phi_{l,r}\right].
\end{equation}
\end{subequations}

Inserting the expansion of $w$ into the definition of $C_k$ we find
\begin{align}
C_k=\frac{1}{2}\sum_{ij} K_{kij} q_i q_j.
\end{align}
Consequently, the equation of motion for the out-of-plane motion becomes

\begin{equation}
\ddot{q}_n + \xi_{0n}^2 q_n +  \sum_{i,j,m} W_{ij;mn} q_i q_j q_m=0,
\end{equation}
with
\begin{equation}
W_{ij;mn}=J_1(\xi_{0n})^2\frac{1}{4}M_{ijmn} - J_1(\xi_{0n})^2\frac{1}{4} \sum_k\frac{J_2(\xi_{1,k})^2}{2\xi_{1,k}^2}K_{kim}K_{kjn}.
\end{equation}
being the effective coupling matrix, which enters Eq. (1) in the main text. From the definition, it is clear that $M_{ijmn}$ is symmetric in all indices. The second term in the definition of the coupling matrix is symmetric under $i\leftrightarrow j$ and $m \leftrightarrow n$, which is reflected in the notation $W_{ij;mn}$. An alternative derivation based on the Airy stress function is outlined in \cite{ermi+13}.
\section*{Magnitude of coupling}
\begin{figure}
		\centering
		\includegraphics[width=1\columnwidth]{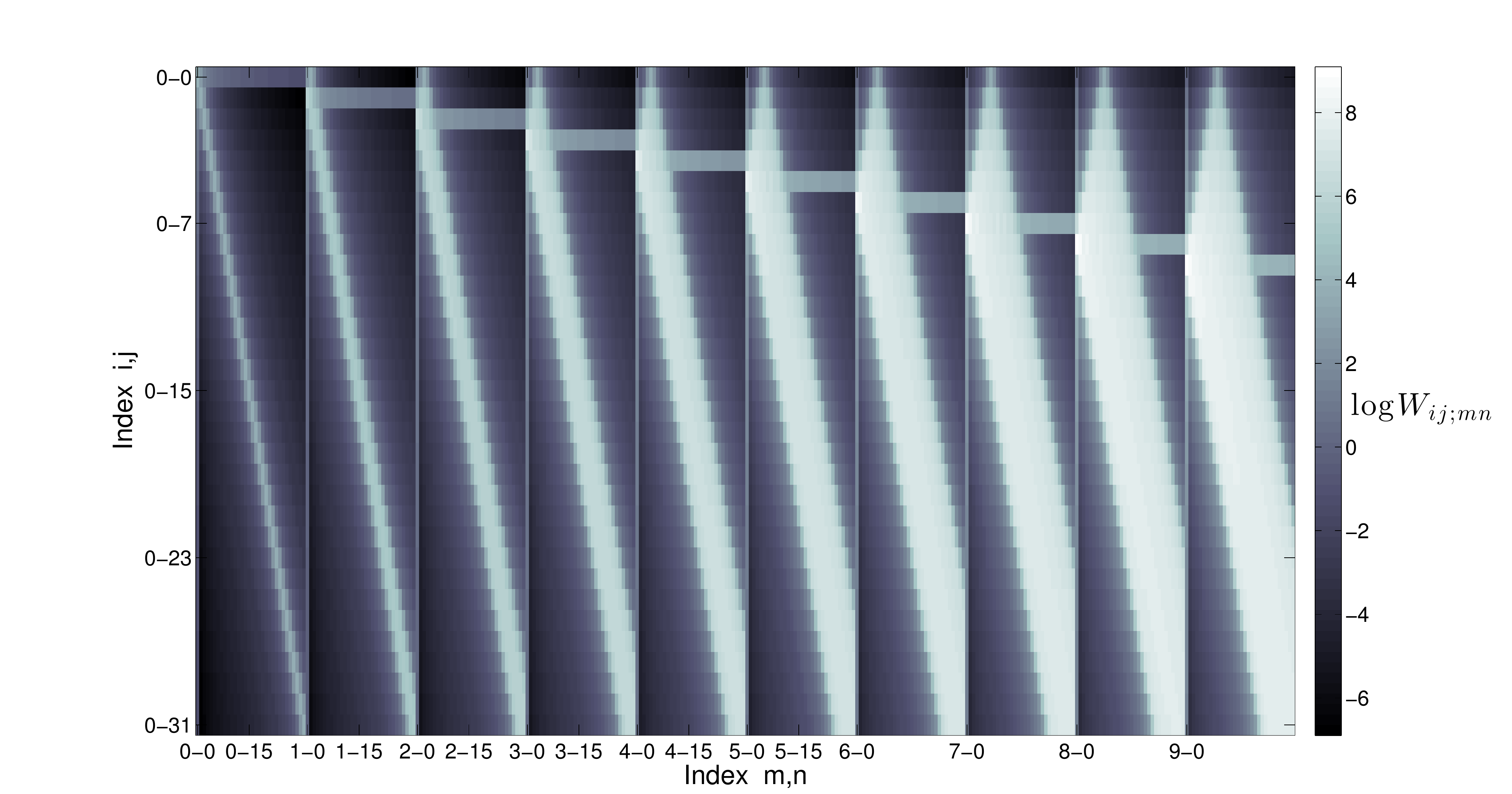}\hfill
				
\caption{Visualization of the nonlinear mode coupling to the fundamental mode. The axes provide the mode number combinations. The coupling generally increases with mode number.\label{VisMatr}}
	\end{figure}

Since we are interested in the dynamics of the fundamental mode, the structure of the matrix can be visualized by fixing the first index to $i=0$, and loop over the remaining indices, treating $ij$ and $mn$ as superindices. From  Fig. \ref{VisMatr}, it is clear that the strength of the coupling as well as the density of modes to which the fundamental mode couples increases with mode number.

	 %can we get this for R>= 9nm?
	
		 %can we get this for R>= 9nm?

\section*{Metastable states in FPU systems}
\begin{figure}
		\centering
		\includegraphics[width=.8\columnwidth]{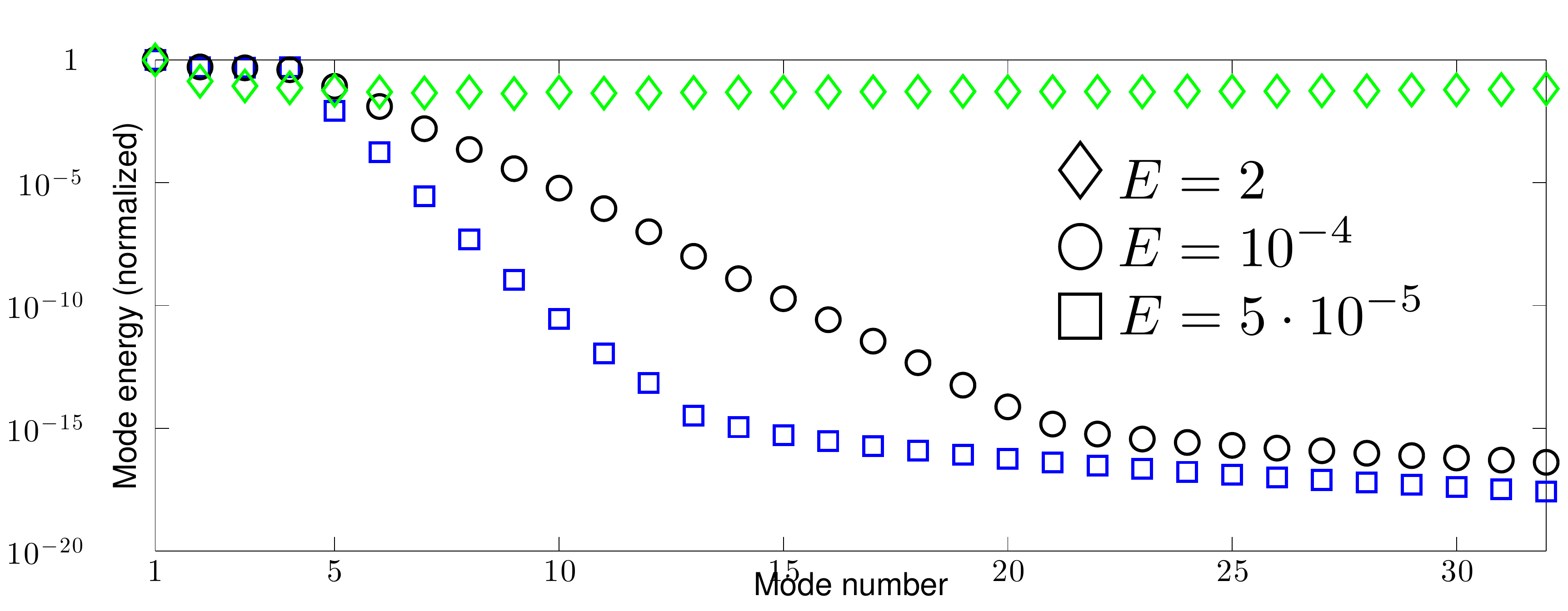}\hfill
				
\caption{Mode energy distribution during the metastable state. The lowest four modes where initialized with equal energy. The energy is exponentially localized in mode space, with an increasing localization length with energy. Blue squares correspond to total energy $E=5\cdot 10^{-5}$, black circles to $E=10^{-4}$ and green diamonds to $E=2$. The mode energies are normalized to the energy of the fundamental mode.\label{fig:Loc}}
\end{figure}

	One of the striking features of the FPU problem is the existence of a long-lived metastable state far from equipartition. An initial understanding of the apparent lack of equipartition was based on the similarity of the FPU model to the Korteweg-de Vries (KdV) equation \cite{zakr+65}. The KdV equation is integrable, and the lack of equipartition was attributed to this fact. However, the metastability persists only for energies below a certain threshold. Above the threshold, the metastable state disappears and the system approaches equipartition on a relatively short time scale \cite{izch+66}. The metastability 
is therefore not completely explained by the similarity of the FPU 
problem to the integrable KdV equation. A complementary approach to describe the metastable 
state was developed by Flach and coworkers \cite{fliv+05}, which focusses on the 
energy spectrum in mode space (q-space). For vanishing nonlinearity,
any excitation consisting of only one mode is a periodic
orbit. For nonzero nonlinearity, the orbit will 
spread out to nearby modes, constituting what is denoted a 
q-breather. These q-breathers are exponentially localized in mode
space, and are obtained by treating the FPU system as a perturbation 
to a system of uncoupled harmonic oscillators. Consequently, the 
q-breather formalism applies to generic nonlinear lattices as long as 
a non-resonance condition is fulfilled and for sufficiently low 
energies. The exponential localization in mode space is present also in the system considered in this paper, see Fig. \ref{fig:Loc}. Here, the four lowest modes are initialized with equal energy. The localization length shows a clear dependence on the excitation energy, and in the limit of large energies the spectrum is essentially flat, signifying equipartition.
	\bibliography{nldmgrex,bibFPU,bibmd}